\documentclass[epj]{svjour}

\newcommand{\be}{\begin{equation}}
\newcommand{\ee}{\end{equation}}
\newcommand{\eqref}[1]{(\ref{#1})}
\newcommand{\text}[1]{\textrm{\scriptsize{#1}}}
\newcommand{\tev}{\, {\rm TeV}}
\newcommand{\gev}{\, {\rm GeV}}

\newcommand{\ord}{{\cal O}}

\usepackage{graphicx}
\usepackage{fancyhdr}
\usepackage{amssymb}

\def\makeheadbox{{
 }}
\def\mytitle{My title} 
\def\myauthors{My name}  
\def\mytype{My type of session}
\def\mysession{My session}

\def\mytitle{Lepton Flavor Violation in the LHT} 
\def\myauthors{Bjoern Duling}    
\def\mytype{Contributed Talk}    
\def\mysession{Alternatives}

\begin{document}
\title{Lepton Flavor Violation in the Littlest Higgs Model with T-Parity}
\subtitle{A Clear Distinction from Supersymmetry}
\author{Bj\"orn Duling
\thanks{\emph{Email:} bduling@ph.tum.de}%
}                     
%
%
\institute{Physik Department, Technische Universit\"at M\"unchen, 85748 Garching, Germany}
%
\date{}
\abstract{
The Littlest Higgs Model with T-Parity (LHT) contains new sources of flavor and CP violation both in the quark and lepton sector. These have their origin in interactions of ordinary fermions with mirror fermions mediated by new heavy gauge bosons. Large deviations from the Standard Model (SM) are to be expected in the lepton sector where tiny neutrino masses suppress the SM predictions by many orders of magnitude below the experimentally accessible level.
Here we give a brief summary of LFV processes relevant for the foreseeable future and point out that correlations between branching ratios of LFV decays in the LHT exhibit a structure vastly different from their analogues in the MSSM, thus allowing for a transparent distinction between these two models. 
%
} 
\maketitle
\section{Introduction}
\label{sec:1}
Little Higgs models \cite{Arkani-Hamed:2001ca,Arkani-Hamed:2001nc,Schmaltz:2005ky,Perelstein:2005ka} offer an alternative route to the solution of the little hierarchy problem. One of the most attractive models of
this class is the Littlest Higgs model~\cite{Arkani-Hamed:2002qy} with T-parity (LHT)~\cite{Cheng:2003ju,Cheng:2004yc},
where a discrete symmetry forbids tree-level corrections to electroweak observables, and thus considerably weakens the constraints coming from electroweak precision data~\cite{Hubisz:2005tx}. {Under this new symmetry the particles have distinct transformation properties, that is, they are either T-even or T-odd. Especially the T-odd particles will have a substantial impact on deviations from the SM. The LHT model is based on a two-stage spontaneous symmetry breaking occurring at the scale $f$ and the electroweak scale $v$. Here the scale $f$ is taken to be larger than about 500\gev, which allows to expand expressions in the small parameter $v/f$. The additionally introduced}
gauge bosons, fermions and scalars are sufficiently light to be 
discovered at LHC and there is a dark matter candidate \cite{Hubisz:2004ft}. Moreover, 
the flavor structure of the LHT model is richer than the one of the 
Standard Model (SM), mainly due to the presence of three doublets 
of mirror quarks and 
leptons and their 
weak interactions with the ordinary quarks and leptons\footnote{A nice roundup of conditions for LFV is given in~\cite{Blum:2007he}.}, as discussed in {\cite{Low:2004xc,Hubisz:2005bd,Blanke:2006xr}}.

In the SM the FCNC processes in the lepton sector, like $\ell_i\to\ell_j\gamma$ and {$\mu\to eee$},
are very strongly suppressed due to tiny neutrino masses. For example the
branching ratio for $\mu\to e \gamma$ in the SM amounts to at most $10^{-54}$,
to be compared with the present experimental upper bound, $1.2\cdot 10^{-11}$ \cite{Brooks:1999pu},
and with the one that will be available within the next two years,
$\sim 10^{-13}$ \cite{Yamada:2005tg,meg:2007wp}. 
Results close to the SM predictions are expected within the LH model without T-parity, where the lepton sector is identical to the one of the SM and the additional $\ord(v^2/f^2)$ corrections have only minor impact on this result. 

A very different situation is to be expected in the LHT model, where the
presence of new flavor violating interactions and of mirror leptons with
masses of order $1\tev$ can  change the SM expectations up to 45
orders of magnitude, bringing the relevant branching ratios for lepton flavor
violating (LFV) processes close to the bounds available presently or in the near future.

\section{LFV in the LHT Model}
\label{sec:2}

\subsection{The Model}
\label{sec:Model}
A detailed description of the LHT model can be found in~{\cite{Blanke:2006eb}, where also a complete set of Feynman rules {has been derived}.} Here we just want to state briefly the
ingredients needed for the {analysis of LFV decays}.

\subsubsection{New Particles}
\label{sec:Particles}
The T-odd gauge boson sector consists of three heavy
``partners'' of the SM gauge bosons,
\begin{equation}\label{2.3}
W_H^\pm\,,\qquad Z_H\,,\qquad A_H\,,
\end{equation}
with masses given to lowest order in $v/f$ by
\begin{equation}\label{2.4}
M_{W_H}=gf\,,\qquad M_{Z_H}=gf\,,\qquad
M_{A_H}=\frac{g'f}{\sqrt{5}}\,.
\end{equation}

The T-even fermion sector contains, in addition to the SM fermions, the heavy top partner $T_+$. 
On the other hand, the T-odd fermion sector~\cite{Low:2004xc} consists of three
generations of mirror quarks and leptons with vectorial
couplings under $SU(2)_L\times U(1)_Y$, that are denoted by
\be
\left(\begin{array}{c}u_H^i\\d_H^i\end{array}\right)\,,\qquad\left(\begin{array}{c}\nu_H^i\\\ell_H^i
\end{array}\right)\qquad(i=1,2,3)\,.
\ee
To first order in $v/f$ the masses of up- and down-type mirror fermions are equal. Naturally, their masses are of order $f$. In the {analysis of LFV decays}, except for $K_{L,S}\to\mu e$, $K_{L,S}\to\pi^0\mu e$, $B_{d,s}\to\ell_i\ell_j$ and $\tau\to\ell\pi,\ell\eta,\ell\eta'$, only  mirror leptons 
{(in contrast to mirror quarks)}
are relevant. 

\subsubsection{Weak Mixing in the Mirror Lepton Sector}
\label{subsec:2.4}
As discussed in detail in~\cite{Hubisz:2005bd}, one of the important ingredients of the mirror sector is the existence of four CKM-like unitary mixing matrices, two for the mirror quarks $(V_{Hu},V_{Hd})$ and two
for the mirror leptons $(V_{H\nu},V_{H\ell})$, that are {related} via
\be\label{2.10}
V_{Hu}^\dagger V_{Hd} = V_\text{CKM}\,,\quad V_{H\nu}^\dagger V_{H\ell} = V_\text{PMNS}^\dagger\,.
\ee
An explicit parameterization of $V_{Hd}$ and $V_{H\ell}$ in terms of three mixing angles and three (non-Majorana) complex  phases can be found in \cite{Blanke:2006xr}.

The mirror mixing matrices  parameterize flavor violating interactions between SM fermions and mirror fermions that are mediated by the heavy gauge bosons $W_H^\pm$, $Z_H$ and $A_H$. The {matrix notation} indicates which of the light fermions of a given electric charge participates in the interaction.

\subsubsection{The Parameters of the LHT Model}
\label{subsec:2.6}

The new parameters of the LHT model, relevant for the study {of LFV decays}, are the symmetry breaking scale $f$ and the mirror lepton masses,
\be
f\,,\quad  m^\ell_{H1}\,,\quad m^\ell_{H2}\,,\quad m^\ell_{H3}\,,
\ee
as well as the mixing parameters of the mirror lepton sector,
\be
\theta_{12}^\ell\,,\quad \theta_{13}^\ell\,,\quad \theta_{23}^\ell\,,\quad \delta_{12}^\ell\,\quad \delta_{13}^\ell\,\quad \delta_{23}^\ell\,.
\ee
{Once the new heavy gauge bosons and mirror fermions will be discovered and their masses measured at the LHC, the only free parameters of the LHT model will be the mixing angles $\theta^\ell_{ij}$ and the complex phases $\delta^\ell_{ij}$ of the matrix $V_{H\ell}$, that can be determined with the help of LFV processes. An analogous set of parameters describing masses and mixings exists in the mirror quark sector and can be probed by FCNC processes in $K$ and $B$ meson systems, as discussed in detail in~\cite{Blanke:2006eb,Blanke:2006sb}.

\subsection{Results}
In~\cite{Blanke:2007db} an extensive analysis of LFV in the LHT model has been given. It includes the decays $\ell_i\to\ell_j\gamma$, $\mu\to eee$, the six three body leptonic decays $\tau^-\to\ell_i^-\ell_j^+\ell_k^-$, the semi-leptonic decays $\tau\to\ell\pi,\ell\eta,\ell\eta'$ and the decays $K_{L,S}\to\mu e$, $K_{L,S}\to\pi^0\mu e$ and $B_{d,s}\to\ell_i\ell_j$ that are flavor violating both in the {quark and lepton sector}. Moreover, $\mu-e$ conversion in nuclei and the flavor conserving $(g-2)_\mu$ have been studied. A detailed phenomenological analysis has been performed in that paper, paying particular attention to various ratios of LFV branching ratios that will be useful for a clear distinction of the LHT model from the MSSM.

In contrast to $K$ and $B$ physics in the LHT model, where the SM {contributions constitute} a sizable and often the  dominant part, the T-even contributions to LFV observables are completely negligible due to the smallness of neutrino masses and the LFV decays considered are entirely governed by mirror fermion contributions. 

\begin{figure*}
\begin{minipage}{0.45\textwidth}
a)\vspace{-.5cm}
\center{
\includegraphics[width=\textwidth]{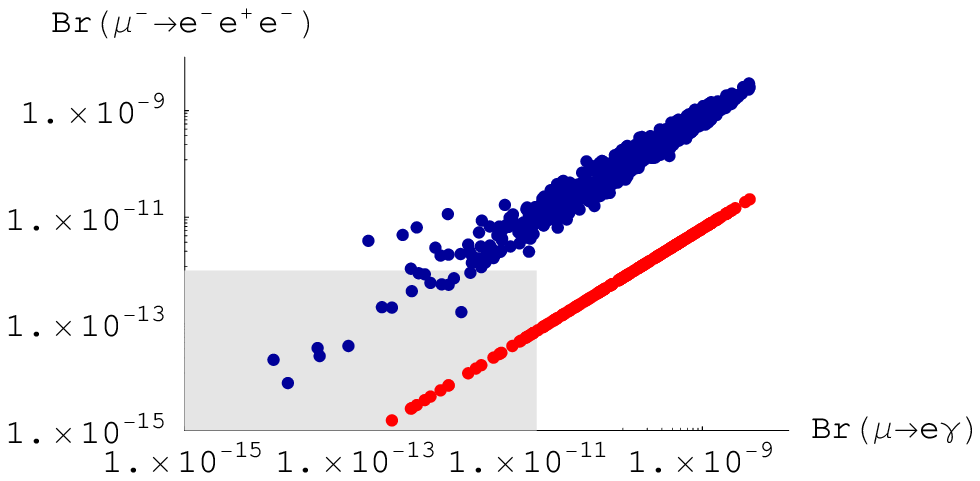}
}
\end{minipage}
\begin{minipage}{0.45\textwidth}
b)\vspace{-.5cm}
\center{
\includegraphics[width=\textwidth]{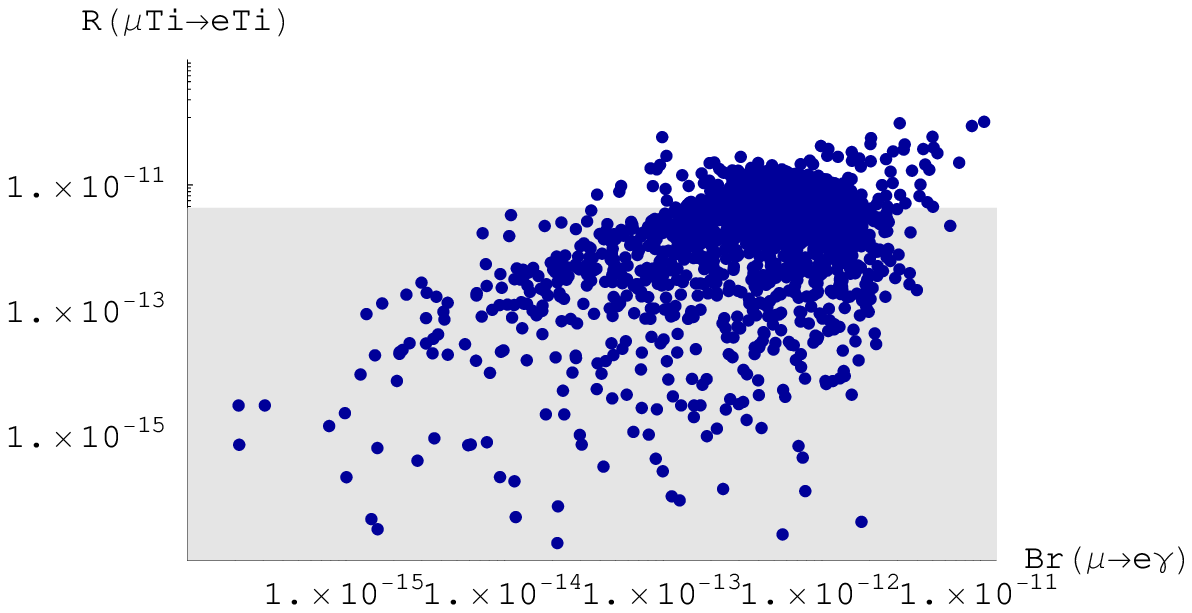}
}
\end{minipage}
\vspace{-0.5cm}
\caption{a) Correlation between $Br(\mu\to e\gamma)$ and $Br(\mu\to eee)$ in the LHT model (upper curve)  \cite{Blanke:2007db}. The lower line represents the dipole contribution to $\mu\to eee$ separately, which{, unlike in the LHT model,} is the dominant contribution in the MSSM. b) $R(\mu\text{Ti}\to e\text{Ti})$ as a function of $Br(\mu\to e\gamma)$, after imposing the existing constraints on $\mu\to e\gamma$ and $\mu\to eee$ \cite{Blanke:2007db}. The present experimental upper bounds are indicated by the shaded areas. \label{fig:meg-m3e}}
\end{figure*}


In order to see how large these contributions can possibly be, it is useful to first consider those decays for which the strongest constraints exist. In this spirit in Fig.~\ref{fig:meg-m3e}.a) we show $Br(\mu\to eee)$ as a function of $Br(\mu\to e\gamma)$, obtained from a general scan over the mirror lepton parameter space, with $f=1\tev$. {It is found that} in order to fulfil the present bounds, either the mirror lepton spectrum has to be quasi-degenerate or the $V_{H\ell}$ matrix must be very hierarchical. Moreover, as shown in Fig.~\ref{fig:meg-m3e}.b), even after imposing the constraints on $\mu\to e\gamma$ and $\mu\to eee$, the $\mu-e$ conversion rate in Ti is very likely to be found close to its current bound, and for some regions of the mirror lepton parameter space even violates this bound.

\begin{table*}
{
\begin{center}
\begin{tabular}{|c|c|c|c|}
\hline
decay & $f=1000\gev$ & $f=500\gev$ & exp.~upper bound \\\hline
$\tau\to e\gamma$ & $8\cdot 10^{-10}$  & $1\cdot 10^{-8}$  & $9.4\cdot10^{-8}$ \cite{Banerjee:2007rj} \\
$\tau\to \mu\gamma$ & $8\cdot 10^{-10}$ &$2\cdot 10^{-8}$  &$1.6\cdot10^{-8}$ \cite{Banerjee:2007rj}\\
$\tau^-\to e^-e^+e^-$ & $7\cdot10^{-10}$ & {$2\cdot10^{-8}$} & $3.6\cdot10^{-8}$ \cite{Abe:2007ev}\\
$\tau^-\to \mu^-\mu^+\mu^-$ & $7\cdot10^{-10}$ & $3\cdot10^{-8}$   & $3.2\cdot10^{-8}$ \cite{Abe:2007ev} \\
$\tau^-\to e^-\mu^+\mu^-$ & $5\cdot10^{-10}$ & $2\cdot10^{-8}$   & $4.1\cdot10^{-8}$ \cite{Abe:2007ev}\\
$\tau^-\to \mu^-e^+e^-$ & $5\cdot10^{-10}$ & $2\cdot10^{-8}$  &$2.7\cdot10^{-8}$ \cite{Abe:2007ev} \\
$\tau^-\to \mu^-e^+\mu^-$ & $5\cdot10^{-14}$  & {$2\cdot10^{-14}$} & $2.3\cdot10^{-8}$ \cite{Abe:2007ev}\\
$\tau^-\to e^-\mu^+e^-$ & $5\cdot10^{-14}$  &{$2\cdot10^{-14}$}   & $2.0\cdot10^{-8}$ \cite{Abe:2007ev} \\
$\tau\to\mu\pi$ & $2\cdot10^{-9} $  & $5.8\cdot10^{-8} $ & $5.8\cdot10^{-8}$ \cite{Banerjee:2007rj}\\
$\tau\to e\pi$ & $2\cdot10^{-9} $ & $4.4\cdot10^{-8} $& $4.4\cdot10^{-8}$ \cite{Banerjee:2007rj}\\
$\tau\to\mu\eta$ & $6\cdot10^{-10}$ & $2\cdot10^{-8}$  &  $5.1\cdot 10^{-8}$ \cite{Banerjee:2007rj}\\
$\tau\to e\eta$ & $6\cdot10^{-10}$  & $2\cdot10^{-8}$  &  $4.5\cdot 10^{-8}$ \cite{Banerjee:2007rj}\\
$\tau\to \mu\eta'$ & $7\cdot10^{-10}$ & $3\cdot10^{-8}$ & $5.3\cdot 10^{-8}$ \cite{Banerjee:2007rj}\\
$\tau\to e\eta'$ & $7\cdot10^{-10}$ & $3\cdot10^{-8}$  & $9.0\cdot 10^{-8}$ \cite{Banerjee:2007rj}\\\hline
\end{tabular}
\end{center}
}
\caption{Upper bounds on LFV $\tau$ decay branching ratios in the LHT model, for two different values of the scale $f$, after imposing the constraints on $\mu\to e\gamma$ and $\mu\to eee$ \cite{Blanke:2007db}. For $f=500\gev$, also the bounds on $\tau\to\mu\pi,e\pi$ have been included. The current experimental upper bounds are also given. {The bounds in \cite{Banerjee:2007rj} have been obtained by combining Belle \cite{Enari:2005gc,Abe:2006sf} and BaBar \cite{Aubert:2006cz,Aubert:2005wa} results.} \label{tab:bounds}}
\end{table*}

The existing constraints on LFV $\tau$ decays are still relatively weak, so that they presently do not provide a useful constraint on the LHT parameter space. However, as seen in Table~\ref{tab:bounds}, most branching ratios in the LHT model can reach the present experimental upper bounds, in particular for low values of $f$, and are very interesting in view of new experiments taking place in this and the coming decade. 

The situation is different in the case of $K_L\to\mu e$, $K_L\to\pi^0\mu e$ and $B_{d,s}\to\ell_i\ell_k$, due to the double GIM suppression in the quark and lepton sectors. $Br(K_L\to\mu e)$ for instance can reach values of at most $3\cdot10^{-13}$ which is still one order of magnitude below the current bound, and  $K_L\to\pi^0\mu e$ is even by two orders of magnitude smaller. Still, measuring the rates for  $K_L\to\mu e$ and $K_L\to\pi^0\mu e$ would be desirable, as
these decays can shed light on the complex phases present in the mirror quark sector.

While the huge enhancements of LFV branching ratios possible in the LHT model are clearly interesting, such effects are common to many other NP models, such as the MSSM, and therefore cannot be used to distinguish these models. However, correlations between various branching ratios should allow a clear distinction of the LHT model from
the MSSM. While in the MSSM \cite{Ellis:2002fe,Arganda:2005ji,Brignole:2004ah,Paradisi:2005tk,Paradisi:2006jp} the {dominant role in decays with three leptons in the final state and in $\mu-e$ conversion in nuclei is typically played} by the dipole operator, in \cite{Blanke:2007db} it is found that this operator is basically irrelevant in the LHT model, where
$Z^0$-penguin and box diagram contributions are {much more important}. As can be seen in Table \ref{tab:ratios} and also in Fig.~\ref{fig:meg-m3e}.a) this implies a striking difference {between} various ratios of branching ratios in the MSSM and in the LHT model. This difference can be made even more transparent by considering double-ratios such as
\be
R=\frac{Br(\tau^-\to e^-e^+e^-)}{Br(\tau^-\to\mu^-\mu^+\mu^-)}\frac{Br(\tau^-\to\mu^-e^+e^-)}{Br(\tau^-\to e^-\mu^+\mu^-)}\,.
\label{eq:double-ratio}
\ee
In the MSSM, where the coefficient of the dominant dipole operator is log-enhanced, the $\mu\leftrightarrow e$ symmetry is strongly broken and \eqref{eq:double-ratio} should considerably deviate from unity. In the LHT however, where the dipole operator is completely negligible, \eqref{eq:double-ratio} is expected to yield unity. Indeed, we find that in the MSSM  $R_\text{MSSM}\simeq 20$ while in the LHT $0.8\lesssim R_\text{LHT}\lesssim 1.2$.
We like to point out that this procedure of comparing ratios and double ratios of branching ratios should be very useful in distinguishing these two models once enough LFV processes have been measured in low energy experiments.
Even if for some decays this distinction is less clear when significant Higgs contributions are present~\cite{Brignole:2004ah,Paradisi:2005tk,Paradisi:2006jp}, it should be easier than through high-energy processes at LHC.

\begin{table*}
{\renewcommand{\arraystretch}{1.25}
\begin{center}
\begin{tabular}{|c|c|c|c|}
\hline
ratio & LHT  & MSSM (dipole) & MSSM (Higgs) \\
\hline
$\frac{Br(\mu^-\to e^-e^+e^-)}{Br(\mu\to e\gamma)}$  & \hspace{.6cm} 0.4\dots2.5\hspace{.6cm}  & $\sim6\cdot10^{-3}$ &$\sim6\cdot10^{-3}$  \\
$\frac{Br(\tau^-\to e^-e^+e^-)}{Br(\tau\to e\gamma)}$   & 0.4\dots2.3     &$\sim1\cdot10^{-2}$ & ${\sim1\cdot10^{-2}}$\\
$\frac{Br(\tau^-\to \mu^-\mu^+\mu^-)}{Br(\tau\to \mu\gamma)}$  &0.4\dots2.3     &$\sim2\cdot10^{-3}$ & $0.06\dots0.1$ \\
$\frac{Br(\tau^-\to e^-\mu^+\mu^-)}{Br(\tau\to e\gamma)}$  & 0.3\dots1.6     &$\sim2\cdot10^{-3}$ & $0.02\dots0.04$ \\
$\frac{Br(\tau^-\to \mu^-e^+e^-)}{Br(\tau\to \mu\gamma)}$  & 0.3\dots1.6    &$\sim1\cdot10^{-2}$ & ${\sim1\cdot10^{-2}}$\\
$\frac{Br(\tau^-\to e^-e^+e^-)}{Br(\tau^-\to e^-\mu^+\mu^-)}$     & 1.3\dots1.7   &$\sim5$ & 0.3\dots0.5\\
$\frac{Br(\tau^-\to \mu^-\mu^+\mu^-)}{Br(\tau^-\to \mu^-e^+e^-)}$   & 1.2\dots1.6    &$\sim0.2$ & 5\dots10 \\
$\frac{R(\mu\text{Ti}\to e\text{Ti})}{Br(\mu\to e\gamma)}$  & $10^{-2}\dots 10^2$     & $\sim 5\cdot 10^{-3}$ & $0.08\dots0.15$ \\
\hline
\end{tabular}
\end{center}
\renewcommand{\arraystretch}{1.0}
}
\caption{Comparison of various ratios of branching ratios in the LHT model and in the MSSM without and with significant Higgs contributions \cite{Blanke:2007db}.\label{tab:ratios}}
\end{table*}
Another possibility to distinguish different models of NP
through LFV processes is given by the measurement of $\mu\to e\gamma$ with polarized muons. Measuring the angular distribution of the outgoing electrons, one can determine the size of left- and right-handed contributions separately \cite{Kuno:1996kv}. In addition, detecting also the electron spin would yield information on the relative phase between these two contributions \cite{Farzan:2007us}. In the case of LFV $\tau$ decays, a Dalitz plot analysis~\cite{Dassinger:2007ru} could also contribute to unravelling the operator structure. We recall that the LHT model is peculiar in respect of not involving any right-handed contributions.

On the other hand, the contribution of mirror leptons to the flavor conserving $(g-2)_\mu$ is negligible~\cite{Blanke:2007db,Choudhury:2006sq}, so that the possible discrepancy between SM prediction and experimental data \cite{Czarnecki:2002nt} can not be cured, in contrast to the MSSM with large $\tan\beta$ and not too heavy scalars.

\enlargethispage{\baselineskip}
\section{Conclusions}
\label{sec:3}
{
We have seen that LFV decays open up an exciting playground for testing the LHT model. Indeed, they could offer a very clear distinction between
this model and supersymmetry. Of particular interest are the ratios  
$Br(\ell_i\to eee)/Br(\ell_i\to e\gamma)$ that are $\ord(1)$ in the LHT model but strongly suppressed in  supersymmetric models even in the presence of
significant Higgs contributions. Similarly, finding the $\mu- e$ conversion rate in nuclei at the same level as $Br(\mu\to e\gamma)$ would point into the direction of LHT physics rather than supersymmetry.
}

\section*{Acknowledgements}
I warmly thank the other authors of~\cite{Blanke:2007db}: M.~Blanke, A.~Buras, A.~Poschenrieder and C.~Tarantino. This work has been supported in part by the Cluster of Excellence ``Origin and Structure of the Universe''. The work of B.D. is supported by GRK 1054 of DFG.


\end{document}